\documentstyle[emulateapj]{article}

\newcommand{\hmpc}{\mbox{ $h^{-1}$ Mpc}}
\newcommand{\etal}{{\rm et~al.}}

\begin{document}

\title{THE LOW-REDSHIFT QUASAR-QUASAR CORRELATION FUNCTION FROM AN
EXTRAGALACTIC H$\alpha$ EMISSION-LINE SURVEY TO z=0.4}

\author{
C.N. Sabbey\altaffilmark{1,2}, 
A. Oemler\altaffilmark{3},
P. Coppi\altaffilmark{1},
C. Baltay\altaffilmark{4},
A. Bongiovanni\altaffilmark{5,6},
G. Bruzual\altaffilmark{5},
C.E. Garcia\altaffilmark{7},
J. Musser\altaffilmark{8},
A.W. Rengstorf\altaffilmark{8},
J.A. Snyder\altaffilmark{4}
}

\begin{center}

\altaffiltext{1}{Yale University, Astronomy Department, 260 Whitney, New
Haven CT 06511}

\altaffiltext{2}{Institute of Astronomy, University of Cambridge, Madingley Road, Cambridge CB3 0HA, UK}

\altaffiltext{3}{Carnegie Observatories, 813 Santa Barbara St., Pasadena CA, 91101}

\altaffiltext{4}{Yale University, Physics Department, P. O. Box 208121, New Haven CT 06520-8121}

\altaffiltext{5}{Centro de Investigaciones de Astronom\'{i}a (CIDA), A. P.
264, M\'{e}rida 5101-A, Venezuela}

\altaffiltext{6}{Universidad Central de Venezuela, Departamento de Fisica, 1042
Caracas, Venezuela}

\altaffiltext{7}{Universidad Complutense de Madrid, Dept.\ de Astrof\'{\i}sica, 
     28040 Madrid, Spain}

\altaffiltext{8}{Indiana University, Dept. of Astronomy, 319 Swain West, Bloomington IN, 47405}

\end{center}

\begin{abstract}

We study the large-scale spatial distribution of low-redshift quasars
and Seyfert~1 galaxies using a sample of 106 luminous emission-line
objects ($\overline{M}_{B} \approx -23$) selected by their H$\alpha$
emission lines in a far-red objective prism survey ($0.2 < z < 0.37$).
Of the 106 objects, 25 were previously known AGN (Veron-Cetty and Veron
2000), and follow-up spectroscopy for an additional 53 objects (including
all object pairs with separation $r < 20 \hmpc$) confirmed 48 AGN and
5 narrow emission-line galaxies (NELGs).  The calculated amplitude of
the spatial two-point correlation function for the emission-line sample
is $A = 0.4 \cdot \overline{\xi}(r < 20 \hmpc) \cdot 20^{1.8} = 142
\pm 53$.  Eliminating the confirmed NELGs from the sample we obtain the
AGN clustering amplitude $A = 98 \pm 54$.  Using Monte Carlo simulations
we reject the hypothesis that the observed pair counts were drawn from
a random distribution at the 99.97\% and 98.6\% confidence levels for
the entire sample and the AGN subset respectively.  We measure a decrease
in the quasar clustering amplitude by a factor of $3.7 \pm 2.0$ between
$z = 0.26$ and $z \approx 1.5$, and present the coordinates, redshifts,
and follow-up spectroscopy for the 15 previously unknown AGN and 4 
luminous NELGs that contribute to the clustering signal.

\end{abstract}


\section{Introduction}

With the emergence of systematic quasar surveys of relatively high
surface density in the early 1980's, the study of the large-scale
spatial distribution of quasars became an active area of research (c.f.\
Osmer 1981).  Currently, there are numerous measurements of the mean
clustering properties of quasars at $1 \lesssim z \lesssim 2$ (e.g.,
Croom and Shanks 1996; La Franca, Andreani and Cristiani 1998), with
improved constraints expected soon from the 2dF quasar survey (Smith
et al. 1996) and the Sloan Digital Sky Survey (Gunn and Weinberg 1995).
However, much less work is available at low redshift ($z \lesssim 0.3$),
for which the dominant UV-excess technique for selecting quasars is less
effective (e.g., Marshall 1985) and relatively small cosmological volumes
are sampled.  Measurement of the clustering of low-redshift AGN, however,
provides the zero point and leverage to discriminate among models for
the evolution of large-scale structure.

The primary studies of AGN clustering at low redshift are that of Boyle
and Mo (1993) and Georgantopoulos and Shanks (1994).  Boyle and Mo
measured the correlation function, $\xi(r)$, for an X-ray sample of 183
AGN ($z < 0.2$) selected in the all-sky {\em Einstein} Extended Medium
Sensitivity Survey.  At small scales they found a marginal detection of
clustering, $\xi(r < 10 \hmpc) = 0.7\pm0.6$.  Georgantopoulos and Shanks
investigated the spatial distribution of a far-infrared sample of 192
Seyfert galaxies (56 Sy1 and 136 Sy2) at $z < 0.1$ selected in the all-sky
IRAS survey.  They detected Seyfert clustering at the $\approx 3\sigma$
confidence level on small scales, with $\xi(r < 20 \hmpc) = 0.52\pm0.13$.
By comparison to measurements of quasar clustering at $\overline{z}
\approx 1.5$, both groups favored a comoving evolution model, in which
$\xi(r)$ is constant with redshift, and marginally excluded a stable
evolution model, $\xi(r) \sim (1+z)^{-1.2}$.

We undertake a similar study using a sample of 106 luminous
emission-line objects ($\approx 95$ Sy1/quasar) identified in a large-area
objective-prism survey (Sabbey 1999; Sabbey \etal\ 2000).  With the unique
far-red bandpass ($6000 < \lambda < 9200$\AA) of the observations,
we are able to select AGN by their H$\alpha$ emission lines to $z
\approx 0.4$ for the first time.  In contrast to the samples described
above, the sample employed here is optically-selected and emphasizes
luminous ($\overline{M}_{B} \approx -23$) Sy1 and quasars at ($0.2 < z <
0.37$), as opposed to lower luminosity Sy2/Sy1 in the nearby universe.
In contrast to searching for blue point sources (UV-excess selection,
Sandage 1965), the H$\alpha$ survey is relatively independent of object
color and morphology, and has a high selection efficiency ($\gtrsim 90$\%)
at bright (and faint) apparent magnitudes, with a large dynamic range
($12 < m_{B} < 20$).

\section{The Low Redshift Quasar Catalog}

The objective-prism data were taken during seven nights in January
through May 1999 with the QUEST 16-CCD driftscan camera on the 1-m
Venezuelan Schmidt telescope (Snyder 1998; Sabbey, Coppi, \& Oemler 1998).
A fully-automated analysis pipeline extracts 1-D spectra and fits a
Gaussian profile to all peaks $> 1\sigma$ above the apparent continuum
to measure emission-line signal-to-noise ratio (SNR), equivalent width
(EW), and center.  The minimum SNR required for selection is a function
of EW, ranging from SNR $> 5.5$ for EW $> 50$\AA\ to SNR $> 2.5$ for EW $>
100$\AA\ (see Sabbey 1999).  The survey covers approximately 700 deg$^{2}$
in the equatorial region and contains 719 emission-line candidates,
of which 11\% are previously known emission-line objects and $< 10$\%
are expected to be false detections.  The magnitude range is $9.7 \leq
m_{B} \leq 20.2$ (see Fig.~1).  Follow-up spectroscopy for a total
of 258 emission-line objects (including 88 below the survey detection
thresholds) confirmed 97 Sy1 and low-$z$ quasars ($z \leq 0.37$), 25
Sy2 ($z \leq 0.49$), 4 quasars ($1.5 \leq z \leq 2.8$) and 132 NELGs.

Of the 258 follow-up spectra, 135 were obtained using the WIYN Hydra
multi-fiber spectrograph (Barden and Armandroff 1995).  During April,
May, and June 1999 a total of 25 Hydra fields were observed (20
deg$^{2}$) to characterize the emission-line sample and establish
target selection criteria.  Fibers were placed on all objects with a
peak above the apparent continuum with SNR $\gtrsim 1.5$.  Of the 71
emission-line candidates in the Hydra sample with SNR $> 2.5$ and EW $>
50$\AA, 67 were confirmed as actual emission-line objects (three of
the four spurious detections were the result of spectrum overlaps in
the objective prism data).  The remaining 123 follow-up spectra were
obtained with a slit spectrograph on the du Pont 2.5m at Las Campanas
(during April 1999 and April 2000) and with FLAIR at the AAO (during
May 2000).  The Las Campanas targets (92 spectra) were preferentially
taken from the AGN candidate list (specified below), while the FLAIR
targets (31 spectra not including duplicate objects or unclassifiable
spectra) were drawn from the full candidate list described above.

The follow-up spectroscopy demonstrated a straightforward selection
criterion for identifying AGN in the sample: of the 78 objects in
the follow-up sample with a candidate emission-line at $\lambda >
7850$\AA\ (i.e., $z>0.2$), 60 were Sy1/quasars, 6 were Sy2, and 12 were
luminous NELGs.  (We identify Sy1/quasars by their broad Balmer lines,
FWHM $> 1000$ km s$^{-1}$, and Sy2 by the line ratio $\lambda 5007 /
\rm{H}_{\beta} > 10$, or $\lambda 5007 / \rm{H}_{\beta} > 2.5$ and $0.5 <
\lambda 6584 / \rm{H}_{\alpha} < 1.5$ (Veilleux and Osterbrock 1987).
The remaining objects are labelled as NELGs.)  In addition, all 116
detections at $\lambda > 7500$\AA\ in the follow-up sample were H$\alpha$,
resulting in extremely reliable H$\alpha$ identification.  The relative
absence of star-forming galaxies at $z > 0.2$ in our survey is expected
due to their faint apparent magnitudes and inverse correlation between
luminosity and emission-line EW (e.g., Fig.~4 in Salzer, MacAlpine,
and Boroson 1989).  In addition, the strong CIV emission of a high
redshift quasar is unlikely to be mistaken for H$\alpha$ due to the
expected appearance of Ly$\alpha$ in our bandpass.

Restricting the emission-line candidates to the detections at $\lambda >
7850$\AA\ ($z>0.2$), with a conservative minimum line SNR $> 2.7$ (EW $>
100$ \AA), yields a sample of 108 AGN candidates.  Based on Veron-Cetty
and Veron (2000) and the NASA Extragalactic Database (NED), 25 of the
108 candidates are previously known objects (21 quasars, 3 Sy1, and
1 Sy2).  The follow-up spectroscopy described above provided spectra
for an additional 55 objects in the AGN candidate list, identifying
45 Sy1/quasar, 3 Sy2, 5 NELGs, and 2 false detections.  The two false
detections were due to spectrum overlaps (reducing the AGN candidate list
to 106 objects), but these had been flagged as possible overlaps based
on the proximity of objects of comparable brightness, and no further
false detections due to overlaps are expected.

The primary selection effect in the objective prism data is the dependence
of the magnitude limit on the emission line equivalent width (e.g.,
Gratton and Osmer 1987).  Of the 10 AGN listed in NED in our $\approx 700$
deg$^{2}$ survey region with $0.2 < z < 0.37$ and USNO $m_{B} < 17.0$,
we independently rediscovered 9 in our AGN candidate list.  For $m_{B} <
18.0$, the fraction rediscovered decreases to 15 out of 25.  All but one
of these not rediscovered in our AGN candidate list have $z \geq 0.3$,
corresponding to the decline in detector quantum efficiency at $\lambda
\gtrsim 8500$\AA.  (If we consider the redshift interval $0.0 < z <
0.37$ then the rediscovery rates in our survey increase to 15 out of
16 for USNO $m_{B} < 17.0$, and 24 out of 36 for USNO $m_{B} < 18.0$).
Calculations of the expected emission-line SNRs for reasonable H$\alpha$
equivalent width distributions gave results comparable to the rediscovery
rates (Sabbey 1999).

A cone diagram of the resulting emission-line catalog of 106 objects
(78 confirmed) is shown in Fig.~2, and the sky coordinates are shown
in Fig.~3.  The objective prism redshift uncertainty is $\sigma_{z}
\approx 0.0042$ (the standard deviation between the prism redshifts and
the 53 follow-up redshifts measured using the [OIII] $\lambda 5007$
line), corresponding to a comoving scale of $\approx 9 \hmpc$ at the
average redshift of the sample $\overline{z} = 0.26$.  The survey volume
is roughly a section of a torus with a radial extent of $\approx 350
\hmpc$ ($\Omega = 1$), a right ascension extent of $> 1000 \hmpc$, and a
declination extent (thickness) of $\approx 70 \hmpc$.  There are 19 pairs
of objects with separation $r < 20 \hmpc$, including one quintuplet and
one triplet of objects.  Follow-up spectroscopy confirming all objects
in pairs except one is shown in Fig.~6, and object coordinates and
redshifts are listed in Table~1.  A paper in preparation (Sabbey et~al.\
2000) describes the survey technique further, providing all follow-up
spectroscopy, coordinates, redshifts, and spectral line measurements.

\section{The Correlation Function}

To quantify the large-scale spatial clustering of the quasar sample, we
measure the two-point quasar-quasar correlation function (Peebles 1980):
\begin{equation}\label{cfeq}
\xi(r) = \frac{N_{\rm obs}(r)}{N_{\rm rand}(r)} - 1,
\end{equation}
where $N_{\rm obs}(r)$ is the observed number of quasar pairs with
comoving separation $r$, and $N_{\rm rand}(r)$ is the average number of
quasar pairs at that separation scale in random comparison catalogs of
the same size.  The comparison catalogs are generated using both
coordinate shuffling and random sampling of the smoothed redshift
distribution (Osmer 1981; Iovino \& Shaver 1988), yielding similar
results (within $0.2 \Delta\xi$).  We calculate the comoving
separations between the quasars within the standard Friedmann model
assuming $H_{0} = 100 \mbox{ $h^{-1}$ km s$^{-1}$ Mpc$^{-1}$}$ and
$\Omega = 1$ (see Kundic 1997).  Qualitatively similar results are
obtained in a nearly empty universe, as expected due to the
independence of {\em relative} object separations on $\Omega_{0}$
(Alcock and Paczy\'{n}ski 1979).


The resulting correlation function is shown in Fig.~4 for the full sample
and with the confirmed NELGs removed from the sample.  On small scales
we detect a marginally significant positive signal (at the $2.7\sigma$
and $1.8\sigma$ levels).  At a scale of $r \sim 30 \hmpc$, $\xi(r)$ drops
below the power law ($\approx 2\sigma$ below), possibly corresponding
to a known feature of galaxy auto-correlation functions (see Peebles
1993, pg.\ 362 and references therein).  At larger scales, $\xi(r)$
is consistent with zero within the uncertainties.  To quantify the
significance of the observed clustering, we use Monte Carlo simulations
to test the null hypothesis that the 19 pairs (12 pairs with the NELGs
removed) were drawn from an unclustered population.  In $10^{6}$ random
catalog simulations produced by shuffling the object redshifts, only
254 simulations for the full sample and 13526 simulations for the AGN
subset produced as many or more pairs at $r < 20 \hmpc$.  Thus, we reject
the hypothesis that the samples do not exhibit clustering on the $r <
20 \hmpc$ scale at the 99.97\% and 98.6\% confidence levels.

To compare the measured AGN clustering strength to that of galaxy systems 
in the local universe and high-redshift quasars (see the following section),
we calculate the AGN clustering amplitude $A$.  The volume-averaged
two-point correlation function on scales $r < R_{0}$ is:
$$
\overline{\xi}(R_{0}) = \frac{1}{V} \int_{V} \xi(r) {\rm d}V = 
  \frac{3}{R_{0}^{3}} \int_{0}^{R_{0}} \xi(r) r^2 {\rm d}r.
$$
Substituting in $\xi(r) = A r^{-1.8}$ we obtain:
$A = 0.4 \overline{\xi}(R_{0}) R_{0}^{1.8}$.  When edge effects are
negligible, $\overline{\xi}(R_{0})$ can be measured by:
$$
\overline{\xi}(R_{0}) = \frac{N_{\rm obs}(r<R_{0})}{N_{\rm rand}(r<R_{0})} - 1,
$$
where $N_{\rm obs}(r<R_{0})$ and $N_{\rm rand}(r<R_{0})$ are the number
of observed and randomly simulated pairs, respectively, with comoving
separations $r < R_{0}$.  We set $R_{0} = 20 \hmpc$, corresponding to our
2$\sigma$ redshift errors, and obtain $A = 98\pm54$ ($N_{\rm obs} = 12$,
$N_{\rm rand} = 5.67$).  Using only objective prism redshifts (i.e., not
using any of the available follow-up redshifts) we obtain $A = 113\pm56$
($N_{\rm obs} = 13$, $N_{\rm rand} = 5.70$).  Removing the one object
in Table~1 that has not been confirmed with follow-up spectroscopy
(and could be a NELG), we obtain $A = 85\pm52$ ($N_{\rm obs} = 11$,
$N_{\rm rand} = 5.60$).

\section{Evolution}

We consider three commonly used, although quite simple, ``models'' for
parameterizing clustering evolution with redshift: 1) comoving evolution, 
$\xi(r) =
{\rm constant}$, 2) stable evolution, $\xi(r) \sim (1+z)^{-1.2}$, and
3) collapsing evolution, $\xi(r) \sim (1+z)^{-3}$.  Thus, we are
assuming evolution of the form:
$$
\xi_{i}(r,z) = A_{i}(z) r^{-1.8}
$$
for a galaxy system $i$.  The dependence of the correlation amplitude on
redshift is given by $A_{i}(z) = A_{i}^{0} (1+z)^{-(1.2 + \epsilon)}$,
where $A_{i}^{0}$ is the clustering amplitude for galaxy system $i$
at $z = 0$, and $\epsilon = -1.2$, 0, and 1.8 in the comoving, stable,
and collapsing models.  We use measured values of the correlation
amplitudes for galaxies, groups, and rich clusters: $A_{\rm Galaxy}^{0}
= 21$ (Peebles 1993), $A_{\rm Group}^{0} = 100$ (Bahcall and Choksi
1991), and $A_{\rm Cluster}^{0} = 360$ (Bahcall and Soneira 1983).
Although significantly weaker correlation amplitudes for rich clusters
have been presented in the literature (e.g., Dalton \etal\ 1992), it has
been suggested that the discrepancy is due to systematic differences in
cluster richness (Bahcall and West 1992).  We therefore use the larger
Bahcall and Soneira (1983) measurement to represent the full range of
clustering strengths observed locally.

The measured AGN amplitude is comparable to the amplitude expected
for clustering of groups of galaxies, but less consistent with the
clustering properties of normal galaxies and rich clusters.  We quantify
this in terms of the observed number of quasar--quasar pairs with
$r < 20 \hmpc$, and the expected number of pairs for each clustering
strength $A_{i}$ in combination with each evolution model.  We first
evaluate the $\xi_{i}(r,z)$ at the average redshift of the observed
quasar sample ($z=0.26$) for the three evolution models, obtaining
$\xi_{i}(r,0.26)$.  Then we integrate $\xi_{i}(r,0.26)$ from $r =
0$ to $20 \hmpc$, obtaining $\overline{\xi}_{i}(20,0.26)$.  Finally,
we multiply $\overline{\xi}_{i}(20,0.26)$ by $N_{\rm rand}(r<20)$ to
obtain the predicted number of quasar--quasar pairs in our sample for each
galaxy system clustering strength.  That is, $N_{\rm predict}^{i}(r<20) =
(1 + \overline{\xi}_{i}(20,0.26)) N_{\rm rand}(r<20)$.  The results are
shown in Table~2.

The number of observed quasar--quasar pairs is consistent with that
predicted for the group clustering strength in each evolution model, but
marginally inconsistent (at the $2\sigma$ level) with galaxy clustering
and our assumed cluster--cluster amplitude.  If confirmed with a larger
sample, this would suggest that $z \sim 0.3$ quasars tend to be located
in small groups of galaxies (groups of $11\pm8$ galaxies based on the
current measurement and Eq.~2a in Bahcall and Choksi 1991).  Indeed,
several imaging studies have indicated that $z \lesssim 0.4$ quasars
tend to reside in small to moderate groups of galaxies (see Hartwick
and Schade 1990 and references therein).  Other studies, however, have
suggested that low-redshift quasars inhabit environments similar to
normal galaxies (Smith, Boyle, and Maddox 1995).

In Fig.~5 we compare the measured quasar clustering amplitude to
measurements at other redshifts.  Currently, it is not even clear
whether the quasar clustering amplitude decreases with redshift (negative
evolution), increases with redshift, or remains constant.  Although a
number of previous studies reported negative evolution (e.g., Iovino
and Shaver 1988; Kruszewski 1988; Iovino, Shaver, and Cristiani 1991),
there have also been reports of constant clustering (e.g., Andreani and
Cristiani 1992) and positive evolution (e.g., La Franca, Andreani, and
Cristiani 1998).  Given the large uncertainties, the current measurement
at $\overline{z} = 0.26$ is consistent with the assumed quasar clustering
amplitude of $A = 27$ at $\overline{z} \approx 1.4$ (La Franca, Andreani,
and Cristiani 1998).

Georgantopoulos and Shanks (GS) measured somewhat weaker clustering
in their low-$z$ Sy1 sample ($A = -9\pm24$), although this possible
discrepancy is only at the $2.0\sigma$ level.  Such a discrepancy could
possibly result from the significantly different samples employed (their
non-optical sample contains lower luminosity, relatively nearby AGN).
For example, IRAS galaxies are known to be relatively biased against
high density regions, and related effects could be manifest in the IRAS
Seyfert sample used by GS.  Also, possible biases in the environments
of low-redshift AGN as a function of AGN luminosity have been suggested
(Fisher \etal\ 1996), and different host galaxies as a function of AGN
luminosity or redshift would be relevant because early-type galaxies
are known to cluster more strongly by a factor of several than late-type
(Davis and Geller 1976; Loveday \etal\ 1995).

This work was supported by the National Science Foundation, the Department
of Energy, and the National Aeronautics and Space Administration.
The Observatorio Astron\'{o}mico Nacional is operated by CIDA for the
Consejo Nacional de Investigaciones Cient\'{\i}ficas y Tecnol\'{o}gicas.
This research has made use of the NASA/IPAC Extragalactic Database (NED)
which is operated by the Jet Propulsion Laboratory, California Institute
of Technology, under contract with the National Aeronautics and Space
Administration.  This research also used the VizieR Catalogue Service
(Ochsenbein et al. 2000).


\clearpage



\begin{figure}
\plotone{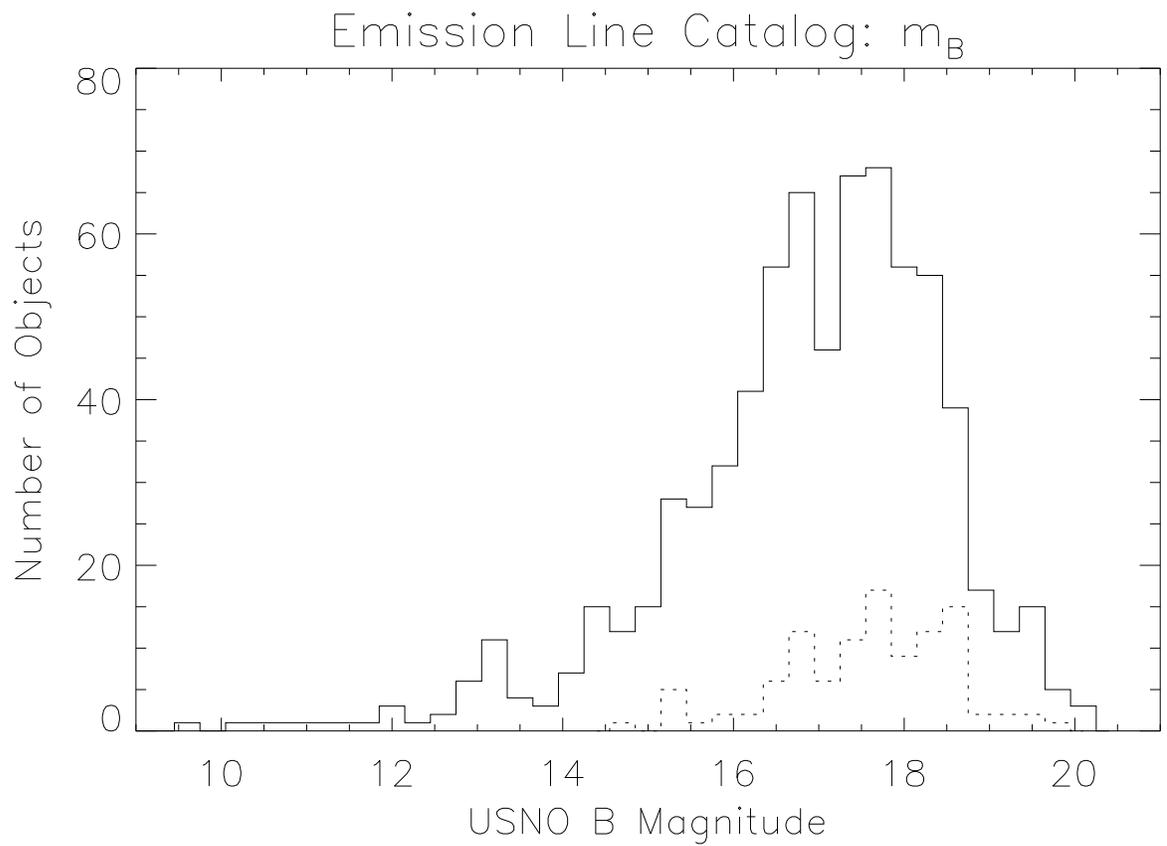}
\figcaption{The histogram of USNO B magnitudes (Monet et al. 1999) is shown 
for the entire emission-line catalog of 719 objects, and the $(0.2 < z < 0.37)$
subset of 106 AGN candidates used in this paper (dashed lines).}
\end{figure}

\begin{figure}
\plotone{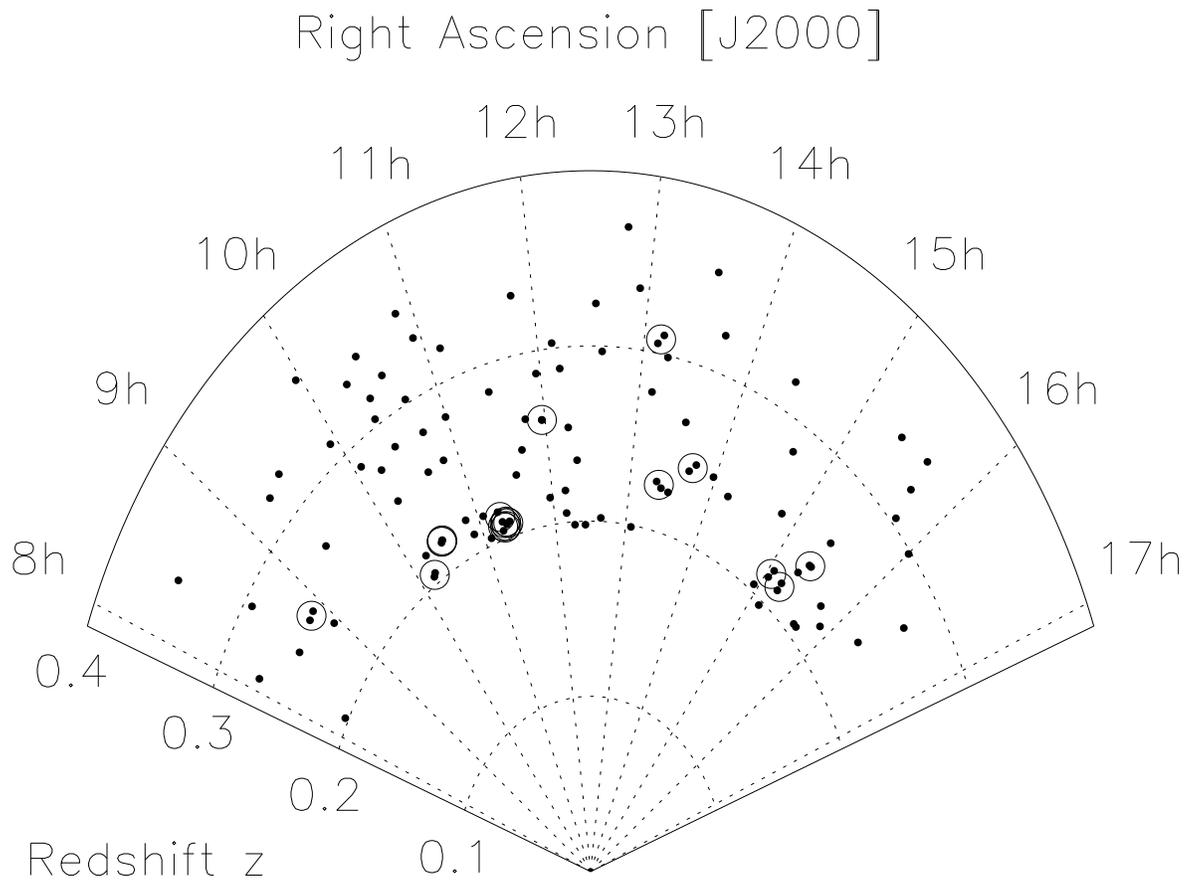}
\figcaption{A cone diagram of the emission-line sample of 106 objects is
shown.  The 19 pairs with comoving separation $r < 20 \hmpc$ are circled.  
\label{cone}}
\end{figure}

\begin{figure}
\plotone{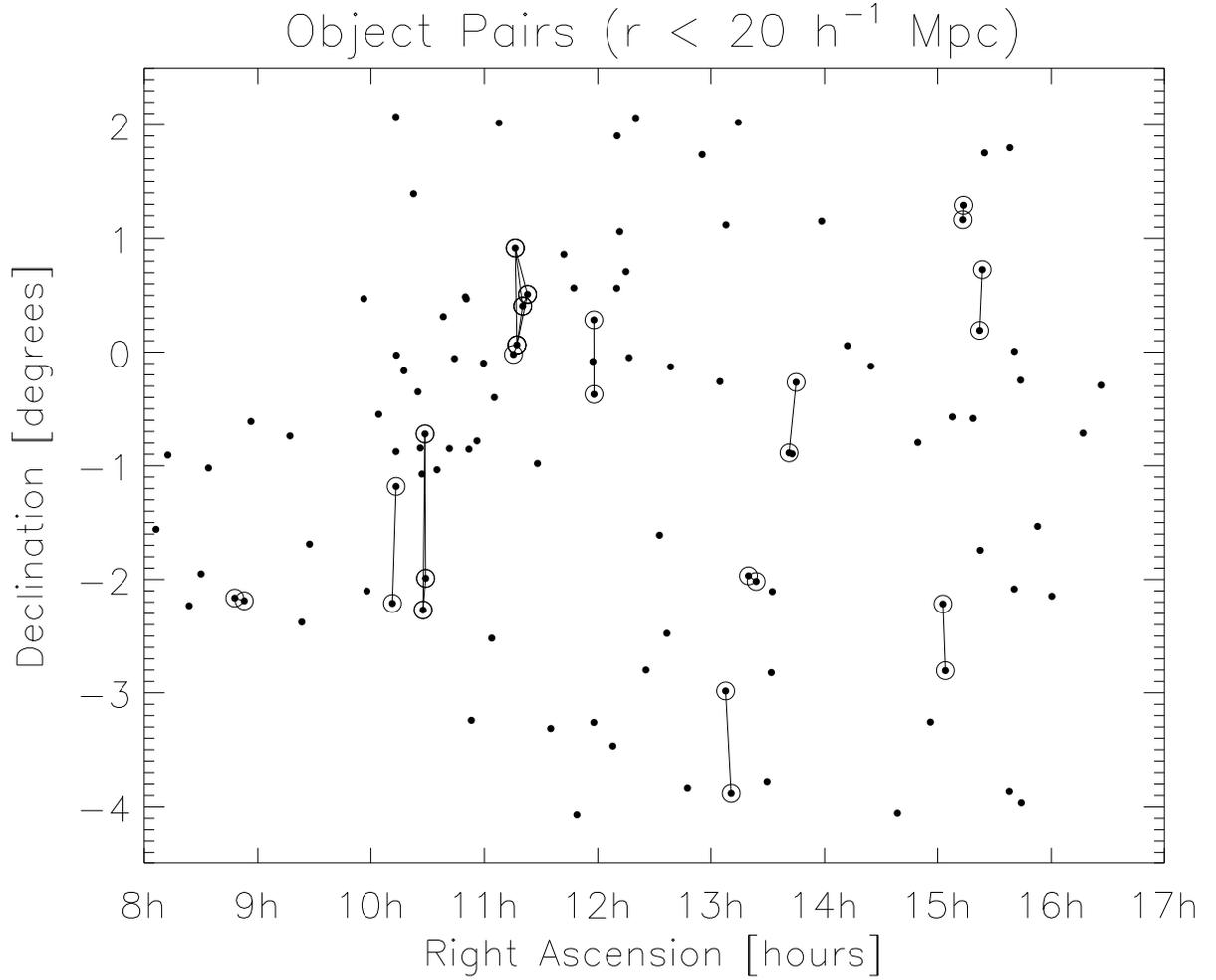}
\figcaption{The sky coordinates of the sample of 106 AGN candidates are shown.
The 19 pairs with comoving separation $r < 20 \hmpc$ are circled.
There is a cluster of five objects near 11h~15m (on a size scale of
$\approx 30 \hmpc$) and a triplet near 10h~30m.  The survey region
($\approx 700$ deg${2}$) extends from about 9h~45m to 16h~15m in the 
declination range $-4.3 < \delta < 2.1$, with extensions to 8h and to 17h 
at $-0.1 < \delta < -2.2$ (the precise boundaries are jagged and a function 
of exposure time due to the driftscan nature of the observations).
\label{pairplot}}
\end{figure}

\begin{figure}
\plotone{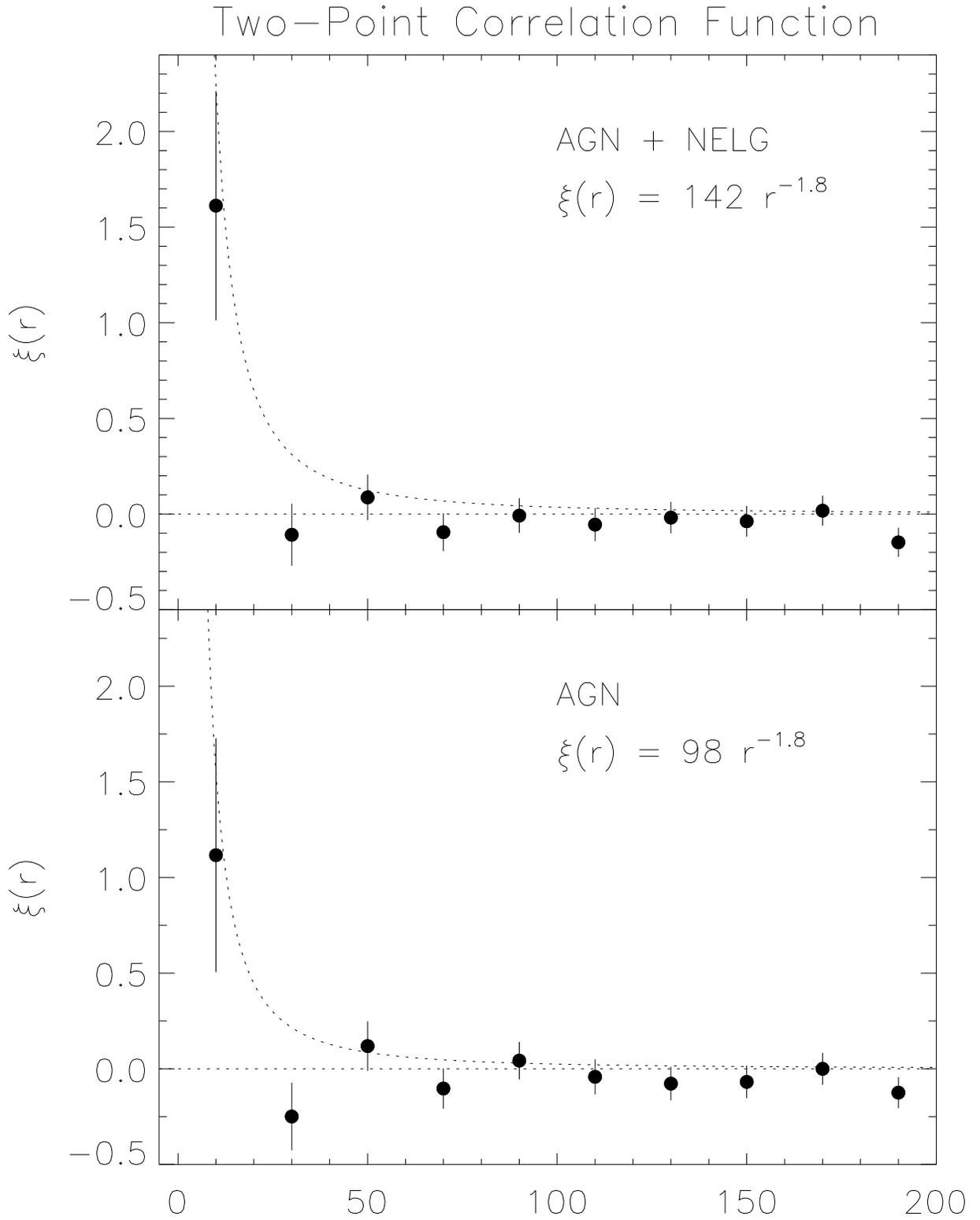}
\figcaption{The $\overline{z} = 0.26$ correlation function
in redshift space with bin sizes of $20 \hmpc$ is shown for the 
emission-line sample (top plot), and the AGN subset (bottom plot).
The uncertainties are from the error estimator 
$\Delta\xi(r) = [(1+\xi(r))/N_{\rm rand}]^{0.5}$.  The $r^{-1.8}$
power laws with amplitude $A = \xi (1 \hmpc) = 142$ and $98$ as calculated in 
the text are overplotted.\label{plotcf}}
\end{figure}

\begin{figure}
\plotone{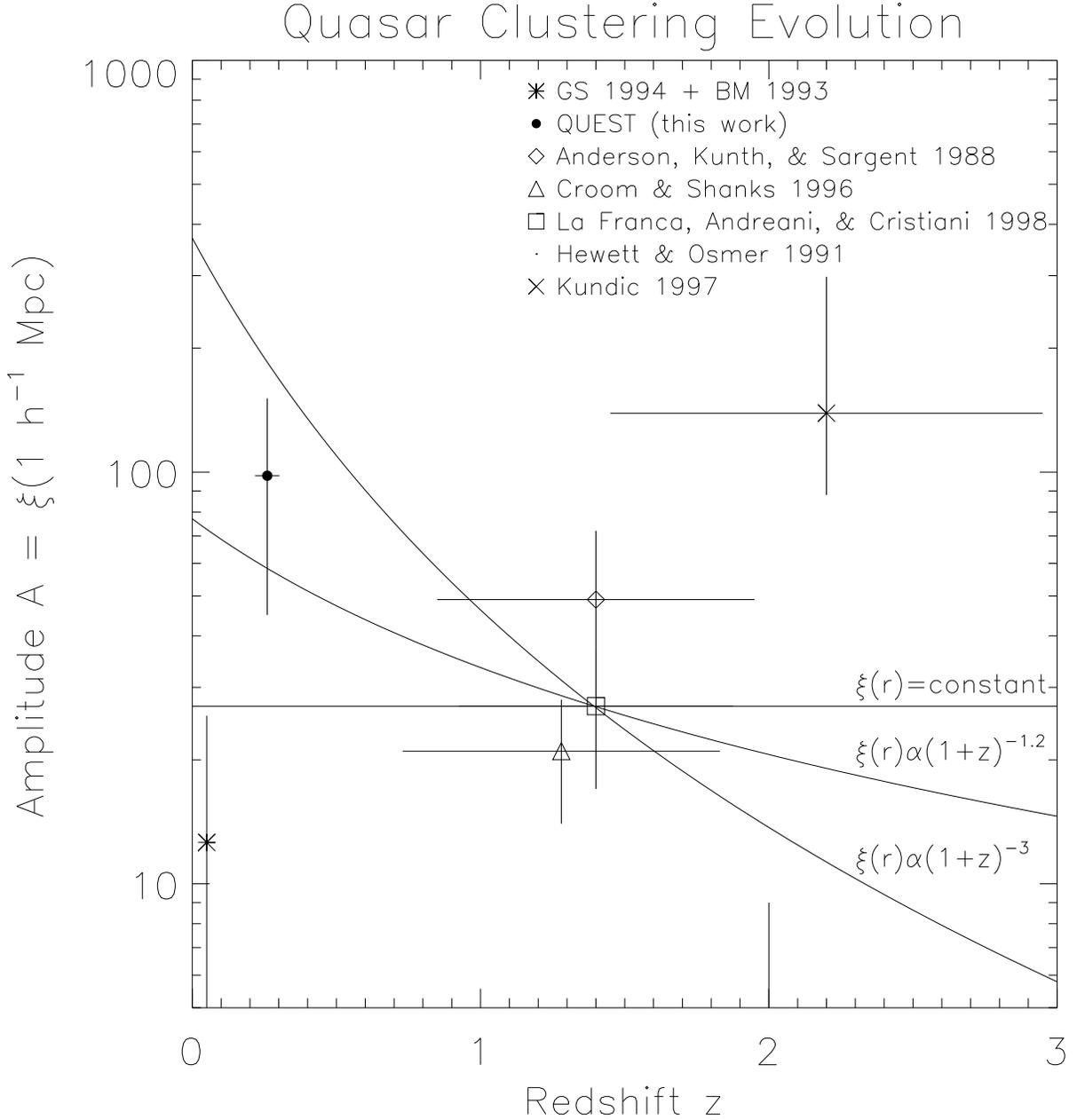}
\figcaption{The measured quasar-quasar correlation amplitude as a
function of redshift based on a representative set of publications.
The full width of the horizontal ``error bars'' indicate 1/2 of the
survey redshift extent.  The $z=0.05$ measurement is from the combined
Georgantopoulos and Shanks (1994) and Boyle and Mo (1993) samples (see
La Franca \etal\ 1998).  For the Sy1 sample of Georgantopoulos and Shanks
(1994), the amplitude is $A = -9\pm24$ (not shown).  The Osmer and Hewett
(1991) measurement at $z=2.0$ is $A = -18\pm27$.  The $\overline{z} =
0.26$ measurement, combined with the previous low-redshift AGN clustering
measurements, is consistent with clustering measurements at $\overline{z}
\approx 1.5$ (suggesting weak or no evolution of quasar clustering out
to intermediate redshift).
\label{cfqso3}}
\end{figure}

\begin{figure}
\plotone{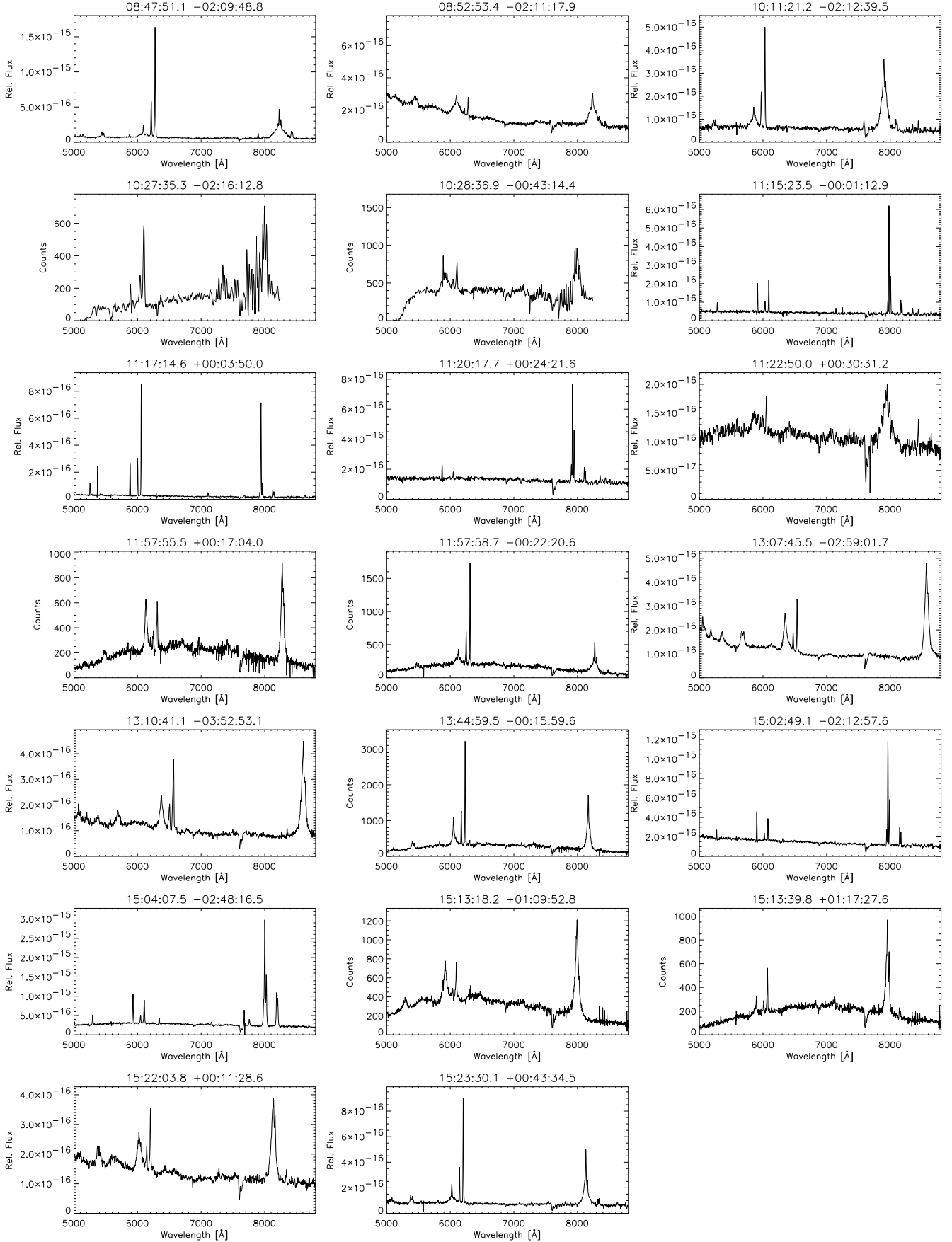}
\figcaption{The follow-up spectroscopy is shown for the emission-line
objects that contribute to the clustering signal on scales $r < 20 \hmpc$
(see Table~1).\label{thespectra}} 
\end{figure}

\clearpage

\begin{deluxetable}{lccllllc}
\small
\tablecaption{Objects contributing to clustering signal ($r < 20 \hmpc$)
\label{pairstable}}
\tablecolumns{8}
\tablewidth{0pt}
\tablehead{ \colhead{Name} & \colhead{R.A. (J2000)} & \colhead{Dec (J2000)} & \colhead{$m_{B}$} & \colhead{$z$} & \colhead{Non-Optical\tablenotemark{3,4}} & \colhead{Sp Class} & \colhead{Ref}  }
\startdata
QUEST0847$-$0209 & 08:47:51.1 & $-$02:09:48.8 & 17.8 & 0.2534 & FIRST/RASS & QSO &   \\ 
QUEST0852$-$0211 & 08:52:53.4 & $-$02:11:17.9 & 16.9 & 0.2545 &            & QSO &   \\ 
QUEST1011$-$0212 & 10:11:21.2 & $-$02:12:39.5 & 18.4 & 0.2045 &            & QSO &   \\ 
Q1010$-$0056  & 10:13:17.2 & $-$01:10:56.1 & 18.4 & 0.202  & RASS       & QSO & 1 \\ 
QUEST1027$-$0216 & 10:27:35.3 & $-$02:16:12.8 & 18.1 & 0.2180 & FIRST      & AGN     &   \\ 
QUEST1028-0043 & 10:28:36.9 & $-$00:43:14.4 & 17.6 & 0.2189 &            & QSO &   \\ 
Q1026$-$0144  & 10:28:57.2 & $-$01:59:22.8 & 16.8 & 0.217  &            & QSO & 1 \\ 
QUEST1115$-$0001 & 11:15:23.5 & $-$00:01:12.9 & 19.7 & 0.2165 &            & NELG    &   \\ 
QUEST1116+0054\tablenotemark{a} & 11:16:19.9 &   +00:54:56.6 & 18.6 & 0.205  &            &         &   \\
QUEST1117+0003 & 11:17:14.6 &   +00:03:50.0 & 19.3 & 0.2101 &            & NELG    &   \\ 
QUEST1120+0024 & 11:20:17.7 &   +00:24:21.6 & 18.6 & 0.2077 & FIRST      & NELG    &   \\ 
QUEST1122+0030 & 11:22:50.0 &   +00:30:31.2 & 17.9 & 0.2088 & RASS       & QSO &   \\ 
QUEST1157+0017 & 11:57:55.5 &   +00:17:04.0 & 18.6 & 0.2603 & RASS       & QSO &   \\ 
QUEST1157$-$0022 & 11:57:58.7 & $-$00:22:20.6 & 17.0 & 0.2603 & RASS       & QSO &   \\ 
QUEST1307$-$0259 & 13:07:45.5 & $-$02:59:01.7 & 17.6 & 0.3056 &            & QSO &   \\ 
QUEST1310$-$0352 & 13:10:41.1 & $-$03:52:53.1 & 17.2 & 0.3109 & RASS       & QSO &   \\ 
Q1317$-$0142  & 13:19:50.4 & $-$01:58:03.5 & 16.8 & 0.225  & FIRST      & QSO & 1 \\ 
Q1321$-$0145  & 13:23:52.8 & $-$02:01:01.7 & 17.5 & 0.224  & FIRST/RASS & QSO & 1 \\ 
UM602       & 13:41:13.9 & $-$00:53:14.8 & 17.4 & 0.237  & FIRST/RASS & QSO & 2 \\ 
Q1342$-$000   & 13:44:59.5 & $-$00:15:59.6 & 17.1 & 0.245  & RASS       & QSO & 2 \\ 
QUEST1502$-$0212 & 15:02:49.1 & $-$02:12:57.6 & 17.4 & 0.2139 & FIRST      & NELG    &   \\
QUEST1504$-$0248 & 15:04:07.5 & $-$02:48:16.5 & 16.6 & 0.2194 & FIRST/RASS & QSO &   \\
QUEST1513+0109 & 15:13:18.2 &   +01:09:52.8 & 17.6 & 0.2174 &            & QSO &   \\ 
QUEST1513+0117 & 15:13:39.8 &   +01:17:27.6 & 18.4 & 0.2123 &            & QSO &   \\ 
QUEST1522+0011 & 15:22:03.8 &   +00:11:28.6 & 17.6 & 0.2389 &            & QSO &   \\ 
QUEST1523+0043 & 15:23:30.1 &   +00:43:34.5 & 17.8 & 0.2391 & RASS       & QSO &   \\
\enddata
\tablerefs{
(1) Hewett, Foltz, and Chaffee 1995;
(2) Surdej et al. 1982;
(3) Voges et al. 1999;
(4) White et al. 1997.
}
\tablenotetext{a}{For this object only a redshift estimate from objective prism data is available.}
\end{deluxetable}

\clearpage

\begin{deluxetable}{ccccccc}
\small
\tablecaption{The observed, random, and predicted quasar-quasar pair counts
with separation $r < 20 \hmpc$ ($\Omega = 1$)
\label{predicttable}}
\tablecolumns{7}
\tablehead{
\colhead{$N_{\rm rand}$} &
\colhead{$N_{\rm obs}$} &
\colhead{$N_{\rm predict}^{\rm Galaxy}$} &
\colhead{$N_{\rm predict}^{\rm Group}$} &
\colhead{$N_{\rm predict}^{\rm \raisebox{0.14em}{\scriptsize Cluster}}$} &
\colhead{$N_{\rm predict}^{\rm QSO}$} &
\colhead{Evolution Model} }
\startdata
\vspace{-1.5em} \\
\phn5.56 & 12 & \phs$7.02$        & \phs12.12\phs          & \phs28.90          & \phs\phn7.42  & Comoving \nl
          &    & {\bf$-1.9\sigma$} &  \phn{\bf\phs$0.0\sigma$}\phs & \phn$+3.1\sigma$ & \phn{\bf $-1.7\sigma$} &        \nl
\tableline \\
\vspace{-2.15em} \\
\phn5.56 & 12 & \phs$6.70$        & \phs10.56\phs      & \phs23.27           & \phs\phn9.44   & Stable \nl
          &    & $-2.1\sigma$ & \phn${\bf-0.4\sigma}$\phs & \phn$+2.3\sigma$ & \phn${\bf-0.8\sigma}$  &  \nl
\tableline \\
\vspace{-2.15em} \\
\phn5.56 & 12 & \phs$6.35$        & \phs\phn8.90\phs     & \phs17.28           & \phs17.71    & Collapsing \nl
          &    & $-2.2\sigma$ & \phn${\bf-1.0\sigma}$\phs & \phn${\bf+1.3\sigma}$ & \phn${\bf+1.4\sigma}$ &     \nl
\vspace{-1.5em} \\
\enddata
\end{deluxetable}


\begin{references}

\reference{a} Alcock, C. \& Paczy\'{n}ski, B., 1979, Nature, 281, 358
\reference{a} Andreani, P., and Cristiani, S. 1992, ApJ, 398, L13
\reference{a} Bagla, J.S. 1998, MNRAS, 297, 251
\reference{a} Bahcall, N.A., and Chokshi, A. 1991, ApJL, 380, L9
\reference{a} Bahcall, N.A., and Soneira, R.M. 1983, ApJ, 270, 20
\reference{a} Bahcall, N.A., and West, M.J. 1992, ApJ, 392, 419
\reference{a} Barden, S.C., and Armandroff, T. 1995, Proc.\ SPIE, 2476, 56
\reference{a} Boyle, B.J., and Mo, H.J. 1993, MNRAS, 260, 925
\reference{a} Croom, S.M., and Shanks, T. 1996, MNRAS, 281, 893
\reference{b} Dalton, G.B., Efstathiou, G., Maddox, S.J., and Sutherland, W.J. 1992, ApJ, 390, L1
\reference{b} Davis, M., Geller, M.J. 1976, ApJ, 208, 13
\reference{b} Loveday, J., Maddox, S.J., Efstathiou, G., Peterson, B.A. 1995, ApJ, 442, 457
\reference{b} Fisher, K.B., Bahcall, J.N., Kirhakos, S., Schneider, D.P. 1996, ApJ, 468, 469
\reference{c} Georgantopoulos, I., and Shanks, T. 1994, MNRAS, 271, 773
\reference{c} Gratton, R.G., \& Osmer, P.S. 1987, PASP, 99, 899
\reference{c} Gunn, J.E. \& Weinberg D.H. 1995, in {\it Wide Field Spectroscopy and the Distant Universe} ed.~S. Maddox \& Ara{g}\`on-Salamanca (World Scientific, Singapore), 3
\reference{c} Hartwick, F.D.A., and Schade, D. 1990, ARA\&A, 28, 437
\reference{a} Hewett, P.C., Foltz, C.B., \& Chaffee F.H. 1995, AJ, 109, 1498
\reference{d} Iovino, A., and Shaver, P.A. 1988, ApJ, 330, L13
\reference{d} Iovino, A., Shaver, P.A., and Cristiani, S. 1991, ASP Conf.\ Ser.\ 21, ed.\ D. Crampton (San Francisco: ASP), 202
\reference{d} Kruszewski, A. 1988, Acta Astronomica, 38, 155
\reference{d} Kundic, T. 1997, ApJ, 482, 631
\reference{e} La Franca, F., Andreani, P., and Cristiani, S. 1998, ApJ, 497, 529
\reference{f} Marshall, H.L. 1985, ApJ, 299, 109
\reference{m} Monet, D., et al. 1996, USNO-SA2.0, (U.S. Naval Observatory, Washington DC)
\reference{m} Ochsenbein F., Bauer P., \& Marcout J., 2000, A\&AS 143, 221 
\reference{g} Osmer, P.S. 1981, ApJ, 274, 762
\reference{g} Peebles, P.J.E. 1980, The Large-Scale Structure of the Universe (Princeton: Princeton Univ.\ Press)
\reference{h} Peebles, P.J.E. 1993, Principles of Physical Cosmology (Princeton: Princeton Univ.\ Press)
\reference{h} Sabbey, C.N. 1999, PhD Thesis, Yale University, ftp://www.astro.yale.edu/pub/sabbey/thesis.ps.gz
\reference{i} Sabbey, C.N., Paolo, C., and Oemler, A. 1998, PASP, 110, 1067
\reference{i} Sabbey, C.N. \etal\ 2000, ApJS, in preparation
\reference{z} Salzer, J.J., MacAlpine, G.M., and Boroson, T.A. 1989, ApJS, 70, 479
\reference{j} Sandage, A. 1965, ApJ, 141, 1560
\reference{k} Shanks, T., Boyle, B.J., and Peterson, B.A. 1988, ASP Conf.\ Ser.\ 2, ed.\ P.\ Osmer and Phillips M.M. (San Francisco: ASP), 244
\reference{k} Smith, R.J. \etal\ 1996, in {\it New Horizons from Multi-Wavelength Sky Surveys}, IAU Symposium 179
\reference{k} Smith, R.J., Boyle, B.J., and Maddox, S.J. 1995, MNRAS, 277, 270
\reference{l} Snyder, J. A. 1998, Proc. SPIE, 3355, 635
\reference{a} Surdej J., Swings J.-P., Arp H., \& Barbier R. 1982, A\&A 114, 182
\reference{a} Veilleux, S., and Osterbrock D.E. 1987, ApJS, 63, 295
\reference{z} Veron-Cetty M.P., Veron P., 2000, ESO Scientific Report 19, 1
\reference{z} Voges et al. 1999, A\&A, 349, 389
\reference{z} White, R.L., Becker, R.H., Helfand, D.J., \& Gregg, M.D. 1997, ApJ, 475, 479
\end{references}
\end{document}